\documentclass[doublecol]{epl2}
\usepackage{graphics}
\usepackage{graphicx}% Include figure files
\usepackage{dcolumn} % Align table columns on decimal point
\usepackage{bm}
\usepackage{color}
\usepackage{epsfig}
\newcommand{\svofa}{Sr$_2$VO$_3$FeAs}
\title{\svofa:A Nanolayered Bimetallic Iron Pnictide Superconductor
}
\author{K.-W. Lee\inst{1} and W. E. Pickett\inst{2}} 
\institute{
 \inst{1}Department of Display and Semiconductor Physics, 
  Korea University, Jochiwon, Chungnam 339-700, Korea\\
 \inst{2}Department of Physics, University of California, Davis,
  CA 95616, USA \\
}
\pacs{74.70.Xa}{Pnictides and chalcogenides}
\pacs{74.20.Pq}{Electronic structure calculations}
\pacs{71.18.+y}{Fermi surface: calculations and measurements; effective mass, 
           g factor}
\abstract{One of the unifying concepts in the iron-pnictide superconductors, 
both for the 
mechanism of magnetic ordering and of unconventional order parameter character, has
been the electron and hole Fermi surfaces that are approximately nested.  
Using the density functional
methods that have predicted Fermi surfaces correctly in SrFe$_2$P$_2$, we find that the 
recently reported 
superconducting \svofa, with $T_c$=37 K and no apparent competition between magnetism
and superconductivity, possesses different Fermi surface geometry
and character than previous classes of iron pnictides.
The intervening layer (a V bilayer) gives rise to bands that cross the Fermi level.
Coupling to the FeAs layer is small except for interaction along the zone boundary,
however that coupling degrades the Fermi surface nesting.
\svofa, with its alternating
layers of open shell atoms, deserves further close study that should 
help to understand the
origin of the properties of iron pnictide compounds.}
\begin{document}
\maketitle

\section{Introduction}
Since Hosono and coworkers found superconductivity with $T_c$=27 K
in iron pnictides,\cite{hosono} the superconducting critical 
temperature $T_c$ has been rapidly increased to 56 K, 
causing excitement in the community second only to the period following
the discovery of high $T_c$ cuprates. So far, five distinct crystal systems
in this class have been uncovered:
${\cal R}$OFeAs (called ``1111"),\cite{cwang,ren}
${\cal M}$Fe$_2$As$_2$ (denoted ``122",
${\cal M}$=Ca, Sr, Ba),\cite{canfield1,tpark,canfield2,gwu,canfield3,sefat}
${\cal A}$FeAs (``111", ${\cal A}$=Li, Na),\cite{tapp,parker,chen}
FeSe$_{1-x}$,\cite{kotegawa,yeh,imai} 
and ``21311'' Sr$_2$${\cal M'}$O$_3$FeAs (also called``42622" since
the unit cell contains two formula unit, ${\cal M'}$=Sc, V).\cite{ogino,xie,zhu}
The superconductivity is expected to be based, broadly speaking,
on the iron coordination and bonding, the band filling, and some 
details that are not yet clear. Some known results are: (1) although there seems to be 
nothing special about the band filling in stoichiometric ${\cal R}$OFeAs,
it requires a small amount of doping (either electron or hole) to
suppress magnetism and allow superconductivity; 
(2) in the 122 class, pressure alone, without doping, drives
the magnetic-to-superconducting change, and in addition
the required pressure is modest ($\sim$5 GPa) by modern standards;
(3) in the ${\cal A}$FeAs, small ${\cal A}$ deficiency results in superconductivity; 
(4) in FeSe$_{1-x}$ either a small amount of Se vacancies or modest pressure 
without doping leads to a high $T_c$ superconductor.

In the 21311 system, Ogino {\it et al.} synthesized superconducting
Sr$_2$ScO$_3$FeP with $T_c$ =17 K.\cite{ogino} 
Xie {\it et al.} prepared isovalent and isostructural Sr$_2$ScO$_3$FeAs.\cite{xie}
In contrast to the 1111 system, in which substitution of P by As leads to enhancing
$T_c$ by $\sim$20 K, this compound is a normal metal, showing Curie-Weiss behavior.
% However, the LDA calculations of Shein and Ivanovskii suggest 
% no remarkable distinction between Sr$_2$ScO$_3$FeP and Sr$_2$ScO$_3$FeAs .\cite{shein}
Very recently, Zhu {\it et al.}, who synthesized superconducting \svofa~ with $T_c$=37 K,
inferred primarily electron-like charge carriers 
through the Hall coefficient measurements and suggested that the strong temperature 
dependence of this coefficient may arise from multiband character.\cite{zhu} 

Most of these Fe pnictides show a clear competition between superconductivity
and magnetism, and the assortment of observed phenomena is leading to
a profusion of ideas about what comprises the essential
characteristics of these superconductors.\cite{mazin,dong,zhiping,yild,qsi,kwlee} 
One of the most widely discussed features is the nesting of two cylinder Fermi
surfaces (FSs) separated by large wavevector Q, being relevant both to magnetic 
ordering and to superconducting order parameter symmetry.
In \svofa, no indication of magnetic order nor even strong temperature dependence of
the susceptibility has been observed, setting it apart from most other Fe pnictide superconductors.
The saturation of the resistivity in high $T$ regime 
\cite{zhu} may represent ``bad metal'' behavior but the large residual resistivity 
suggests sample imperfections should be kept in mind.

This \svofa~ member brings new, distinctive features into 
consideration in Fe-based superconductors. 
First, a bilayer of open shell V ions (presumably close to $d^2$ formal configuration that would leave the FeAs layer with a formal -1 charge as in other classes) 
lies between the FeAs layers, with centers separated by lattice constant $c$=15.7 \AA,
whereas other members have only closed shell ions in this layer.
Secondly, \svofa~superconducts at high 
temperature (37 K) without
requiring {\it either} explicit doping or pressure.
In the various classes of Fe-based superconductors, the degree of
two-dimensionality and its importance has been much discussed. In \svofa,~
the FeAs layers 
are very weakly coupled through conducting V bilayers, although 
(we will show) the one identifiable
coupling is an important one.

Below we present the electronic structure and FSs, which show strong differences
compared to other Fe pnictides, and discuss implications
for the \svofa~ system and its superconductivity, using first principles density 
functional theory. The FSs from such calculations have recently been verified by 
de Haas-van Alphen measurements in SrFe$_2$P$_2$.\cite{analytis}
A study of the oxygen concentration dependence of the properties of \svofa~ indicate
T$_c$ is maximum (near 40 K) for the stoichiometric composition,\cite{FeiHan} so 
study of the stoichiometric compound is most relevant. 
The effects of the open-shell V bilayer on the properties must be addressed in tandem 
with the metallic Fe layer.
The lack of any reported magnetic behavior\cite{zhu,FeiHan} is an important clue.  
As in other Fe-pnictides, the
tendency toward magnetic behavior is overemphasized with the local density approximation.
The types of (overestimated or unphysical) magnetic solutions that can be 
obtained will be left to a separate report.

\begin{figure}[tbp]
%\vskip 8mm
\centerline{{\resizebox{4.0cm}{7cm}{\includegraphics{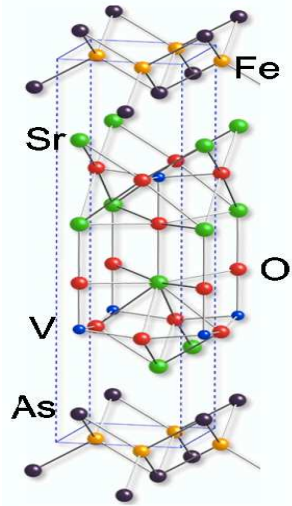}}}}
%{\resizebox{5.0cm}{8cm}{\includegraphics{svofa_str.ps}}}
\caption{(Color online) Crystal structure of \svofa.
The FeAs layers are separated by two perovskite SrVO$_3$ layers with
incomplete O octahedra.
The O ions are not coplanar with Sr or, respectively, V positions.
}
\label{str}
\end{figure}

\section{Structure and Calculation}
In the tetragonal unit cell (space group: $P4/nmm$, No. 129),\cite{zhu} 
displayed in Fig. \ref{str}, 
Fe atoms lie at $2a$ sites ($\frac{1}{4}$,$\frac{3}{4}$,0),
V, Sr, As, and O atoms lie at $2c$ sites ($\frac{1}{4}$,$\frac{1}{4}$,$z$),
and another O atom sits at the $4f$ sites ($\frac{1}{4}$,$\frac{3}{4}$,$z$).
In our calculations, the experimentally observed lattice parameters
$a$=3.9296 and $c$=15.6732~\AA~ were used.\cite{zhu}
The experiment values of the internal parameters $z$ also were
used: 0.3081 for V, 0.8097 and 0.5855 for Sr sites, 0.0909 for As,
0.4318 for O at $2c$ sites, and 0.2922 for O at $4f$ sites. 
The Fe plane
is sandwiched between two V planes (with As between) at a distance of 4.83 \AA,
and the separation between V layers 
(in the $c$ direction) is 6.0 \AA.  These large interplanar 
distances suggest very small dispersion perpendicular to the layers, as will be
shown from the calculations.

In this structure, the V ion is coordinated by five oxygens and one As ion.
The bond lengths of V-O are 1.94 and 1.98 \AA~ in $a$-$b$ plane and 
$c$-direction, respectively, while the V-As separation is 3.40 \AA~and 
breaks the local symmetry of the V $t_{2g}$ orbitals.  In the characterization scheme
of Eschrig and Koepernik,\cite{eschrig} the Fe-Fe distance $d$ = 2.78 \AA~and the 
ratio $r$ of the distance of the As layers from the Fe layer to $d$ is $r$=0.511.
Both of these values lie nearly exactly between the small $d$, 
large $r$, values of FeSe and LiFeAs,
and the high $d$, small $r$, values of LaOFeAs and BaFe$_2$As$_2$.  Noteworthy is that
the $r$ values lies closer to the ideal As tetrahedron value of $r$=0.50
than in the other classes.  

The accurate all-electron full-potential local orbital code, FPLO-7,\cite{fplo1} 
was used for all calculations. The Brillouin zone was sampled with a regular
fine mesh up to 405 irreducible $k$ points, necessary for sampling
the FSs adequately.  

%\section{Results}
\section{Electronic Structure and Fermi Surface}
Shein and Ivanovskii have reported\cite{shein} a ferromagnetic 
electronic structure.  We have obtained several magnetic solutions, differing
by the relative orientation of Fe and V moments.
However, since
no sign of magnetism has been seen in the data, and LDA calculations also overestimate
the tendency toward magnetism in several other Fe pnictide compounds, we focus here
on the nonmagnetic solution.  Even in the magnetic Fe pnictides, it is the electronic 
structure of the non-magnetic system that has received the most intense attention.

\begin{figure}[tbp]
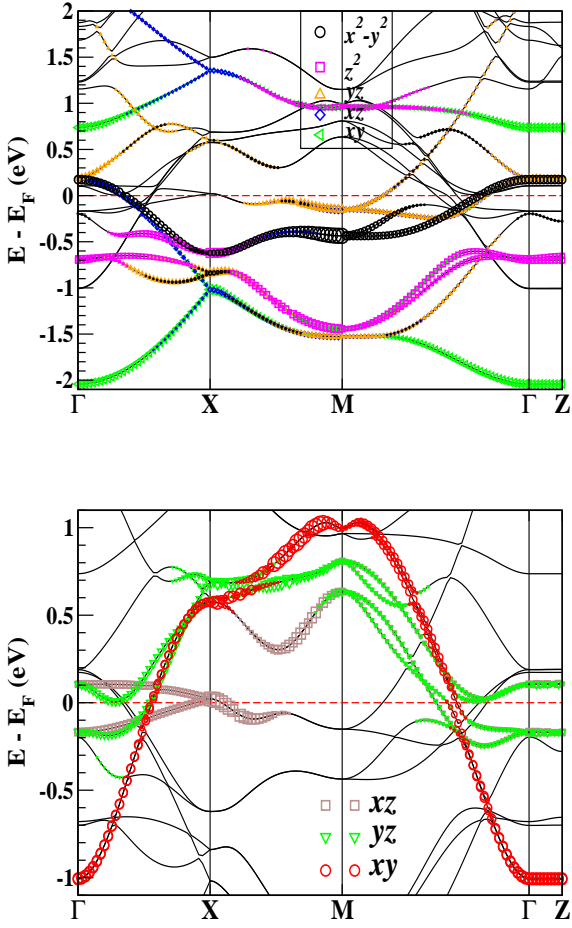

\vskip 8mm
\centerline{{\resizebox{7.5cm}{5.5cm}{\includegraphics{Fig2a.eps}}}}
\vskip 12mm
\centerline{{\resizebox{7.5cm}{5.5cm}{\includegraphics{Fig2b.eps}}}}
\caption{(Color online) Band structure of nonmagnetic \svofa. 
Top: Fe fatbands emphasis in the --2 to 2 eV range. 
Bottom: fatband representation of V $t_{2g}$ states.
The symbol size is proportional to fractional character of each orbital.
The horizontal (red) dashed lines indicate the Fermi energy $E_F$,
set to zero. Note that each band is either Fe or V, with the only
substantial mixing occurring along the X-M line at E$_F$ (see text).
}
\label{pmband}
\end{figure}

\begin{figure}[tbp]
\vskip 8mm
\centerline{{\resizebox{7.5cm}{5.5cm}{\includegraphics{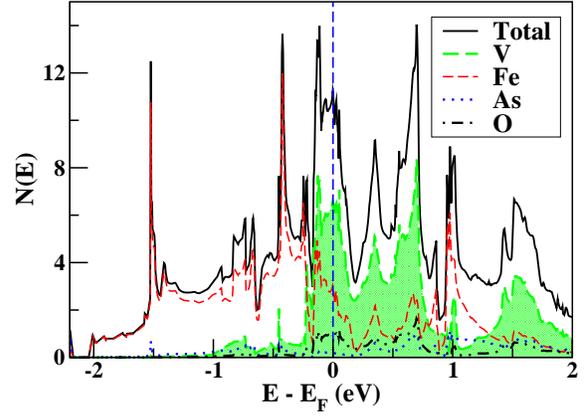}}}}
\caption{(Color online) Total and atom-projected
densities of states (DOSs) near $E_F$ for nonmagnetic \svofa.
The $t_{2g}$-$e_{g}$ crystal field splitting of V is about 1.3 eV.
DOS at $E_F$ $N(E_F)$ = 11.2 states/eV per Fe, composed roughly of
60\% V, 25\% Fe, and 10\% from O ions.   
The Fermi energy is the zero of energy.
}
\label{pmdos}
\end{figure}

\begin{figure}[tbp]
%\flushleft
\centerline{\resizebox{5cm}{4.5cm}{\includegraphics{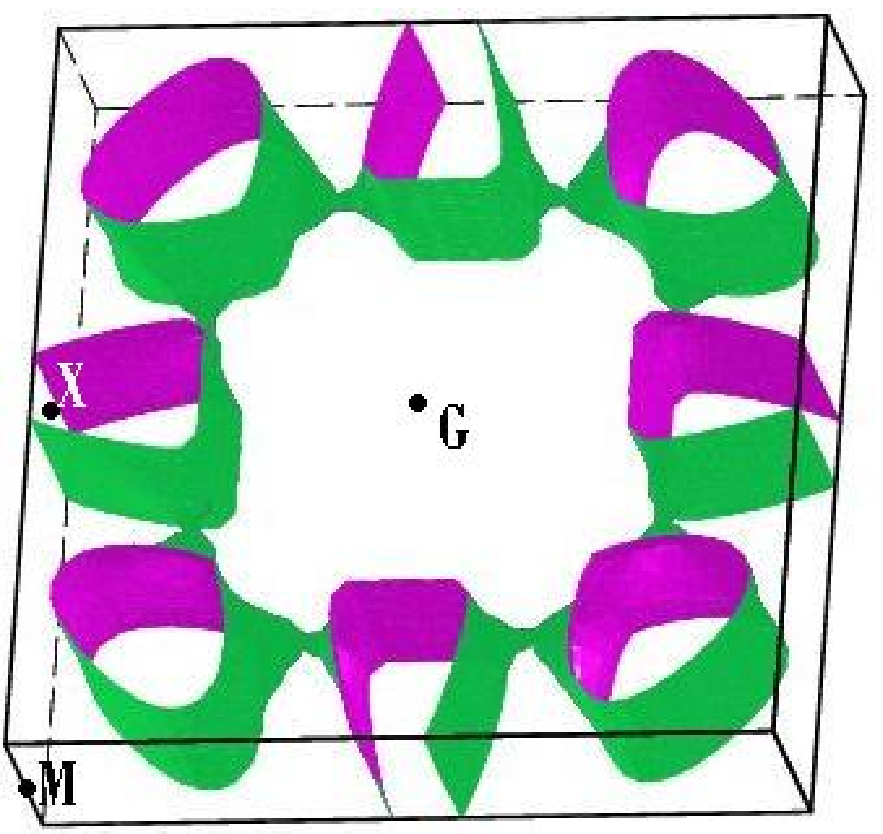}}}
%\flushright
%\vskip -29mm
\centerline{\resizebox{5cm}{4.5cm}{\includegraphics{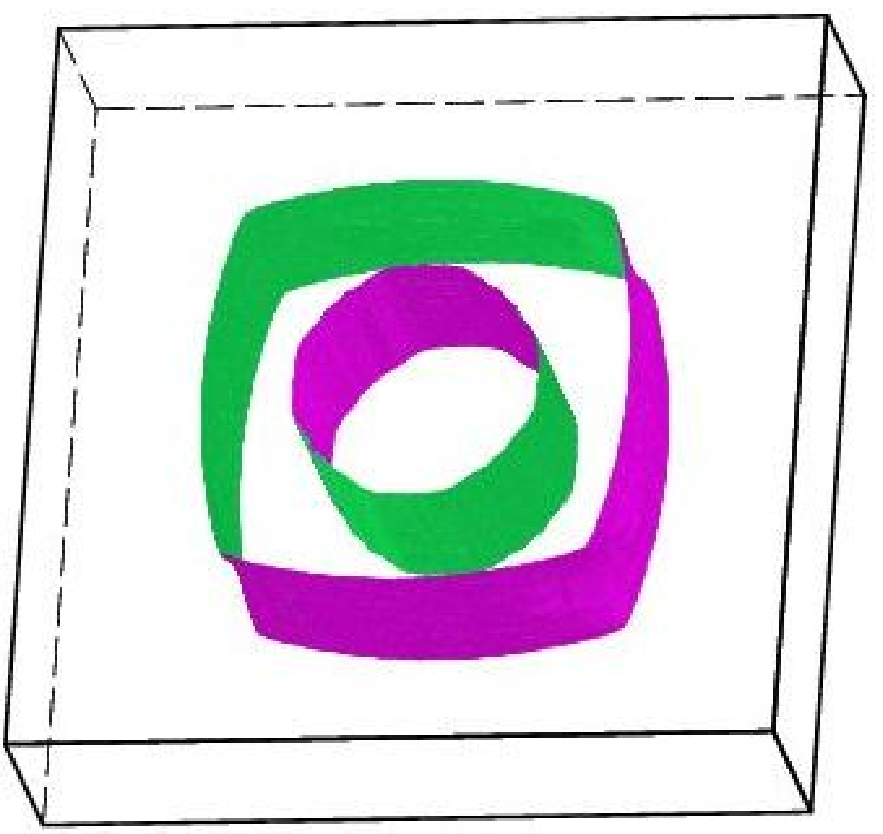}}}
%\flushright
%\vskip -29mm
\centerline{\resizebox{5cm}{4.5cm}{\includegraphics{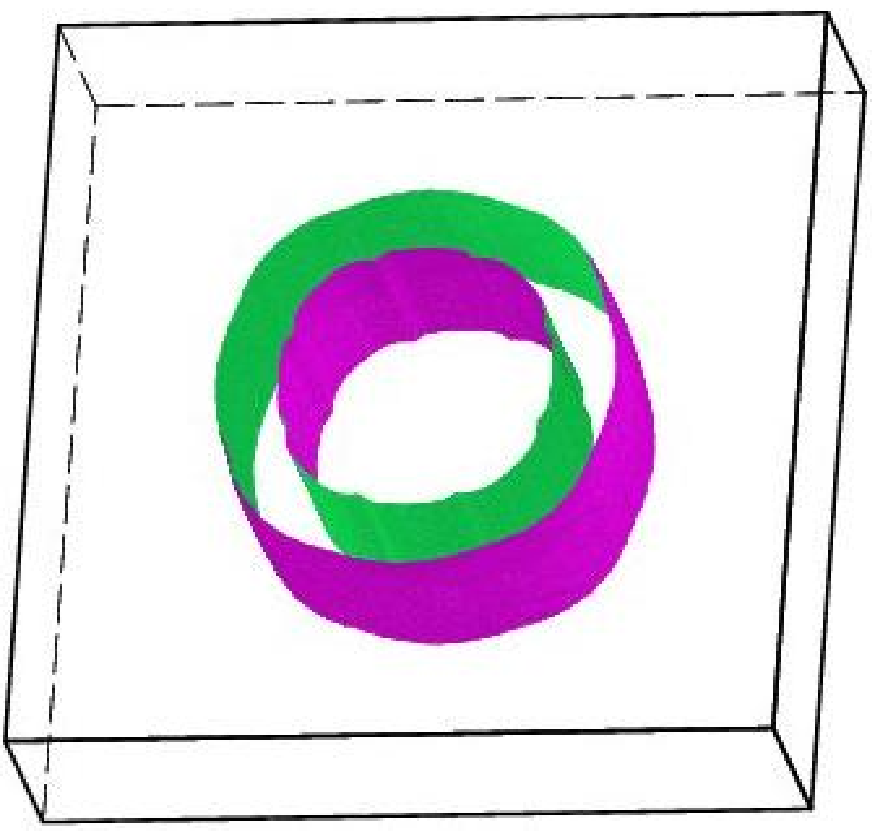}}}
\caption{(Color online) Fermi surfaces of \svofa, showing the
 two-dimensionality.
 Top panel: $X$-centered V hole bowties and rounded triangles lying
along the (1,1) directions. G denotes the $\Gamma$ point.
 Middle and bottom: larger surfaces in both panels are V character and
 contain electrons; smaller cylinders enclose Fe-derived holes.
 Some tiny surfaces are not shown.
}
\label{FS}
\end{figure}

The band structure within 2 eV of the Fermi level (E$_F$),
which is the regime of most of the Fe and V $3d$ character,
is shown in  Fig. \ref{pmband}. 
The bands crossing E$_F$, and in fact throughout the
$d$ band complex, are easily identifiable as either V- or Fe-derived. With one
exception (see below) there
is little mixing of both characters into any single band, reflecting the weak
electronic coupling between the V and Fe subsystems. 

The most distinctive characteristic, compared with other 
iron pnictides,\cite{mazin08,fma08} 
is the presence of {\it metallic} V $t_{2g}$ bands in the ``spacer layer''
and these bands overlap the Fe $3d$ bands.  The projected density of states in 
Fig. \ref{pmdos} indicates that, while a large majority of the V $3d$ states
are unfilled, there is a substantial contribution at the Fermi level and for several tenths of
eV below.
The V $d_{xz}, d_{yz}$ bands are centered (Fig. \ref{pmband}) near 0.3 eV with a bandwidth
of 1 eV, while the $d_{xy}$ band is centered at E$_F$ ({\it i.e.} is half-filled)
with twice larger bandwidth 2 eV.  This $d_{xy}$ band can be represented by 
an independent tight-binding band with near-neighbor hopping amplitude 
$t$=0.24 eV and second
neighbor $t'$=0.07 eV.  Without mixing to other bands, this $d_{xy}$ band would 
give rise to two nearly circular and nearly degenerate Fermi surfaces.
Note that the V $d_{xy}$ band is doubly degenerate 
corresponding to the two V layers, for which the $d_{xy}$ orbitals are 
uncoupled to high precision (the splitting is not visible in the plot).  
The minor difference in band centers of the symmetry-split
V $t_{2g}$ bands reflects the non-zero but small
V site tetragonal crystal field.
There are also four $d_{xz}, d_{yz}$ bands because of the two V ions; 
these are not degenerate but are split by up to 0.5 eV in various parts of
the zone, indicating some coupling of these
orbitals between the  V layers or to the FeAs layers.  

The half-filled $d_{xy}$ band contains about one electron (probably less), while the two partially
filled bands contribute roughly another electron, for each of the two V ions in
the unit cell.   However, since band characters are mixed, these are only
rough estimates.  The projected DOS of Fig. 3 provide a better guide.  The
V $t_{2g}$ DOS lies in the range -0.3 to 1 eV, and seems to be rather less
than 1/3 filled, which would indicate less than two $3d$ electrons per V ion.
If this is indeed true (and it is not possible to divide the charge in an
unambiguous way), then the FeAs layer would be electron-doped in stoichiometric
\svofa~compared to other undoped FeAs compounds.  Then no extrinsic doping 
would be required to give a high T$_c$ superconducting state, as seems to be
the case experimentally.

The Fermi surfaces are presented in Fig. \ref{FS}.
The fatbands representation of Fig. 2 indicate a Fe $d_{x^2-y^2}$ band less than 1 eV
wide straddling E$_F$ and giving rise to one hole FS centered at $\Gamma$, 
while both $d_{xy}$ and $d_{z^2}$ bands have
negligible character at or very near E$_F$.  There is strong Fe $d_{xz}, d_{yz}$ character
at E$_F$, giving rise to the smallest (also hole) FS 
cylinder centered at the $\Gamma$ point.

The other FSs (hole bowties centered at X, a hole rounded-triangle FS lying along
the (110) direction, and a squarish electron FS centered at $\Gamma$)
provide Fermi surfaces not seen in previous Fe-pnictide classes.  The triangle FSs
contain Fe $xy, yz$ character along the $\Gamma$-M line, and in other Fe pnictides this
character dominates the electron cylinder at M. However, as a result of the only mixing 
of V and Fe bands that is visible
along symmetry lines, a hybridization gap of 0.4 eV opens up along the X-M direction
so no M-centered cylinders emerge and the approximate nesting observed in other classes 
is substantially degraded.
LDA calculations reproduce the FSs reliably in the nonmagnetic phase, as shown
recently for SrFe$_2$P$_2$.\cite{analytis} 
% This lack of nesting, which requires the cylindrical FS at M, poses a problem
% for pairing mechanisms or order parameter symmetries that rely on nesting of FSs 
% separated by large Q. 

The difference in electronic structure compared to other FeAs classes has been 
addressed by Mazin,\cite{mazin2} who graphically showed the change in the M-centered,
Fe-derived Fermi surfaces.  He found not only degraded nesting but also a strong 
minimum in the (matrix-element-less)    
generalized susceptibility $\chi_{\circ}(q)$.  However, if the processes contributing
to $\chi_{\circ}(q)$ are confined to Fe (i.e. excluding V), then $\chi_{\circ}^{Fe}(q)$
has a broad maximum around the M point similar to other FeAs compounds, even without
Fermi surface nesting.  The magnetic behavior within the Fe layer should be modeled more
realistically by $\chi_{\circ}^{Fe}(q)$  than by $\chi_{\circ}(q)$, and can be
expected to be much the same in \svofa~as in other FeAs compounds.  Full matrix 
elements should be included to understand the susceptibility more completely; when
included in LaFeAsO, the peak around the M point remains but is reduced in
magnitude.\cite{Monni} 

The contribution of Fe ions to the DOS at $E_F$ $N(E_F)$ is about 50\% larger 
than in other iron pnictides, due partially to the different band 
filling but probably also to the different angular structure of the FeAs$_4$ tetrahedron
discussed in the introduction.  
Note also that 60\% of the total $N(E_F)$ arises from 
the roughly one-third filled $t_{2g}$ manifold of V bands.

The Fe FS cylinders consist of two cylinders with
radii of 0.28$\frac{\pi}{a}$ and 0.35$\frac{\pi}{a}$, containing 0.50 and 0.76 holes
respectively.
The two Fe-based cylinder FSs are in fact much like the important surfaces 
in MgB$_2$: two
similarly sized concentric cylinders at the zone center.  The generalized
susceptibility of such FSs is known,\cite{mgb2}: there 
is structure at 2$k_F$, and the ``nesting function'' has a weak divergence at
2$k_F$. 
One notable feature of such FSs 
is that for boson exchange scattering through
wavevector $q$, the strength is distributed over the entire
region $|q| < 2k_F$, which here comprises more than a third of the area of the zone.  
% A particular inference is that there is no reason
% to expect an $s^{\pm}$ symmetry of superconducting order parameter ({\it i. e.},
% a different sign on the two FSs) in \svofa.

\section{Discussion}
Now we collect our findings and discuss implications.  The metallic V bilayer separating
FeAs layers is a feature not present in other Fe pnictide classes.  This primary
distinction leads to several secondary differences.  The coupling is
weak except between V $d_{xz}, d_{yz}$ to Fe $d_{xz}, d_{yz}$ along the 
zone boundary X-M line, 
but this coupling seriously degrades the approximate
nesting that is seen in other Fe pnictides.  More study is needed to determine
whether this change of nesting accounts for 
the lack of magnetic ordering.\cite{zhu} Also important is that this coupling
is sufficient to establish a common Fermi level, so that an integer number of
electrons is not required in either subsystem, and the Fe band filling deviates somewhat
from related materials.  Superconductivity in
the Fe layer will induce a non-zero order parameter in the V layers via
the proximity effect, but a likely result is that the gaps  in the two layers will be
very different in size, leading to multi-gap (``multiband'') superconducting character. 
The strong differences between \svofa~and other Fe pnictides should be of importance
in identifying the pairing mechanism of these impressive high temperature
superconductors.

The broader consequences for \svofa~that can be expected 
due to the metallic V-O layer -- a new conduction channel
between the Fe layers,  altered band filling and
Fermi surface, etc. -- deserve comment.  Because \svofa~
superconducts at 37 K, in the same temperature range as
several other Fe pnictides, it seems highly likely that its
superconducting mechanism is the same.  
The question of gap symmetry becomes 
more interesting as the Fermi surface develops more sheets
and has strong variation of (Fe, V) character, and thereby
of magnitude of gap.  The gap could be extremely weak on the
V sheets (although magnetic fluctuation may be strong in that
layer as well), leading to an extreme two-gap scenario in addition
to possible phase changes between different surfaces.

The occurrence of a periodic arrangement of strongly
superconducting and weakly superconducting (proximity-induced)
conducting layers within 5 \AA~of each other may be one of the more unusual aspects
of \svofa, altering other properties if not T$_c$.  For example,
a modest magnetic field might drive the V layer normal
while the Fe layer remains strongly superconducting, leading
to unexpected transport and spectroscopic behaviors.
We expect \svofa~to display a number of properties, in the normal
state as well as in the superconducting, that are
different from the previous Fe pnictides, and which should help to clarify 
the microscopic mechanisms underlying the remarkable properties of 
the Fe pnictides.

Note added.  Since the submission of this paper, Wang {\it et al.}\cite{wang} have
reported the electronic structure of \svofa.  That work uses theoretical
atomic positions rather than the experimental ones, and differs in some
respects from those presented here and by Mazin.\cite{mazin2}

\section{Acknowledgments}
We acknowledge helpful communication with I. I. Mazin, R. Weht, and H.-H. Wen.
This work was supported by DOE under Grant No. DE-FG02-04ER46111, and interaction
within DOE's Computational Materials Science Network is acknowledged.
K.W.L. was partially supported by a Korea University Grant No. K0718021.

\end{document}